\documentclass[aps,nofootinbib,showpacs,twocolumn,longbibliography]{revtex4-1}
\usepackage{amstext,amsmath,amssymb,amsfonts,bbm}
\usepackage[latin1]{inputenc}
\usepackage{graphicx}
\usepackage{hyperref}
\usepackage{epstopdf}
\usepackage{amsthm}
\usepackage{tocvsec2}
\usepackage{enumerate}
\usepackage{subfigure}
\usepackage{color}

\def\beq{\begin{equation}}
\def\be{\begin{equation}}
\def\ee{\end{equation}}
\def\bes{\begin{eqnarray}}
\def\ees{\end{eqnarray}}

\theoremstyle{definition}
\theoremstyle{definition}
\theoremstyle{definition}
\theoremstyle{definition}
\theoremstyle{definition}
\theoremstyle{definition}

\begin{document}
\maxtocdepth{subsection}

\title{\large \bf Discretizing parametrized systems: the magic of \emph{Ditt}--invariance}
\author{{\bf Carlo Rovelli}}
\affiliation{Centre de Physique Th\'eorique, Case 907,  Luminy, F-13288 Marseille, EU}

\date{\small\today}

\begin{abstract}\noindent
Peculiar phenomena  appear in the discretization of a system invariant under reparametrization.  The structure of the continuum limit is markedly  different from the usual one, as in lattice QCD. First, the continuum limit does \emph{not} require tuning a parameter in the action to a critical value. Rather, there is a regime where the system approaches a sort of asymptotic topological invariance (``\emph{Ditt}--invariance"). Second, in this regime the expansion in the number of discretization points provides a good approximation to the  transition amplitudes.  These phenomena are relevant for understanding the continuum limit of quantum gravity. I illustrate them here in the context of a  simple system. 

\end{abstract} 

\maketitle


\section{Introduction} 

Discretization plays an important  role in the analysis of many physical system, and can even be used to \emph{define} the theory, as in the lattice definition of QCD  (see \cite{Creutz:1985fk}). A number of characteristic phenomena appear when discretizing a continuum theory. For instance, (i) the continuum theory is recovered taking suitable parameters to their critical values; and (ii) energy conservation is broken by the discretization, and recovered only in the limit.

Discretization plays an important role also in quantum gravity, where loop-quantum-gravity  spinfoam transition amplitudes on fixed foams (see for instance  \cite{Rovelli:2010vv, Rovelli:2011eq}) have been shown to be strictly related to a Regge-like discretization of general relativity \cite{Conrady:2008ea,Barrett:2009mw,Bianchi:2009ri,Rovelli:2011kf,Magliaro:2011qm,Magliaro:2011dz}. However, conventional wisdom about discretization does not appear to apply in the gravitational context, and the structure of the continuum limit appears to be intriguingly different from the conventional one.

To elucidate the situation and investigate the source of the difference, I study here the discretization of a  simple system (discussed in \cite{Bahr:2011uj}) that shares  with general relativity the property of being ``\emph{Diff}--invariant", that is, invariant reparametrization of its evolution parameter.  I show that the discretization of this system contradicts conventional wisdom on discretization and displays the same peculiar features that appear in loop  gravity.  

First, energy is conserved in the discretization. The invariance of the system is gauge-fixed in the discretization \cite{Marsden:2001uq,Bahr:2009ku}, by an additional independent equation, absent in the continuum (In the Appendix, I discuss the sense in which this is a breaking of \emph{Diff}--invariance). This equation fixes the time steps in such a way to conserve energy.

Second, and most importantly, the continuum limit does \emph{not} require the system to go to a critical point, as is the case for normal systems. The continuum limit is simply given by taking the number of discretization points to infinity, \emph{without tuning any parameter}.  This behavior is very surprising at first, given the common behavior of systems under discretization, but it a simple consequence of the scaling structure of the  theory: the parameter in which the discretization is taken is dimensionless. 

Third, correlation functions in parameter time are meaningless, because of the invariance. Physical quantities can instead be derived from transition amplitudes, which are functions of the boundary values of the path integral.  The values of the boundary values spans different regimes for the system.

In particular, there exists a regime where the curvature of the classical trajectory determined by the boundary values is small. In such ``flatness" regime, the classical trajectories are nearly free and the system displays a very remarkable behavior: \emph{the amplitudes become independent from the number of discretization points} \cite{Bahr:2009mc}. In quantum gravity parlance, the system approaches a ``topological" phase. 

Fourth, in approaching this regime, \emph{Diff}--invariance reappears in the discretized theory. In the vicinity of the limit, the system displays an ``almost invariance" \cite{Rocek:1982fr,Hartle:1985wr,Morse:1991te,Hamber:1996pj,Freidel:2002dw}.  This peculiar  ``almost  \emph{Diff}--invariance" and its importance for quantum gravity has been studied by Bianca Dittrich and collaborators \cite{Dittrich:2008pw,Bahr:2009qc,Bahr:2009mc,Bahr:2010cq,Bahr:2011uj}, and I denote it here \emph{Ditt}--invariance (from ``Dittrich").    

Finally, the consequence of \emph{Ditt}--invariance on the discretized path integral, combined with the absence of parameters to be tuned in the continuum limit, is remarkable: in this regime, a very rough discretization yields near exact results.  I show this numerically, below.   In this regime, \emph{the number of discretization points provides  a good perturbation expansion for the amplitudes}.  In a sense, for a parametrized system the flatness regime relates discretization with perturbation theory. 

These findings provide some ground for the hypothesis that the same structure of the continuum limit might work in quantum gravity. In particular, they suggests that: (i) The continuum limit is recovered by refining the foam, without taking any parameter to a critical value. (ii) There is a regime, where the system approaches \emph{Ditt}--invariance, where a perturbation expansion in the number of discretization cells is viable. (iii) In quantum gravity this regime is near flatness. It is also tempting to speculate that the underlying topological theory be BF theory.  This will be discussed more in detail in the conclusions.

\section{Discretization} 

Consider a harmonic oscillator with mass $m$ and angular frequency $\omega$. The action is 
\be
  S=\frac{m}{2} \int dt \left(\left(\frac{dq}{dt}\right)^2-\omega^2 q^2 \right).
\ee
Choose a fixed time interval $t$ of interest and divide it into a large number $N$ of small steps of size $a=t/N$. The continuum theory will be recovered with $N\to\infty$ and $a\to 0$, keeping the size $t=Na$ of the time interval fixed.  Discretize the system on the time steps $t_n=an$, with integer $n=1,...,N$. The system is then described by the variables $q_n=q(t_n)$ and the action can be discretized as follows 
\be
  S_N=\frac{m}{2} \sum_n a \left(\left(\frac{q_{n+1}-q_{n}}{a}\right)^2-\omega^2 q_n^2 \right).
\label{discrete1}
\ee
A standard lattice procedure is then to define  rescaled dimensionless variables $Q_n=\sqrt{\frac{m}{a \hbar} }\,q_n$ and $\Omega=a\omega$, 
so that the dimensionless action becomes
\be
 \frac{S_N}\hbar=\frac{1}2 \sum_n \left((Q_{n+1}-Q_{n})^2-\Omega^2 Q_n^2 \right) \equiv  S_{N,\Omega}(Q_n)
\ee
where all quantities on the r.h.s.\;are now dimensionless.  This action (or better, its analytical continuation in Euclidean time) can be studied numerically to give an approximation of the path integral
\be
 \int D[q(t)]\ e^{\frac{i}{\hbar}{S[q(t)]} } \to  \int dQ_n\ e^{i S_{N,\Omega}(Q_n)}
 \ee
To take the continuum limit we have to send $N\to\infty$, but this is not sufficient: we must also send $\Omega$ to its critical value $\Omega= 0$, since $\Omega=a\omega$ and $a$ must go to zero in the limit. More precisely, say we want to compute the propagator
\begin{multline}
W(q_f,t_f;q_i,t_i)=\langle q_f|e^{-\frac{i}{\hbar}H(t_f-t_i)}|q_i\rangle\\
= \int_{q(t_i)=q_i}^{q(t_f)=q_f} D[q(t)]\ e^{\frac{i}{\hbar}{S[q(t)]}}.
\label{prop}
 \end{multline}
Then this is given by 
\be
W(q_f,t_f;q_i,t_i)= \lim_{\substack{\Omega\to0 \\  N\to\infty}} {\cal N}  \int dQ_n\ e^{iS_{N,\Omega}(Q_n)}
 \ee
where $Q_0=\sqrt{\frac{m}{a \hbar} }q_i$ and $Q_{N}=\sqrt{\frac{m}{a \hbar} }q_f$ and $\cal N$ is a suitable normalization factor for the measure.  

The fact that the limit is obtained not only by taking $N\to\infty$ but also sending $\Omega$ to its critical value is an essential defining feature of the discretization. Near the critical value the correlation lengths of the discretized system diverge in the number of lattice steps, so that they remain finite in physical separations. Taking the discretized systems to its critical point  can implement universality and wash away the effect of the details of the discretization. This same behavior is present in field theory.  It is natural  to think that this pattern is universal. But things are different when discretizing reparametrization-invariant systems. 

\section{Parametrization} 

Consider the system defined by the two variables $q(\tau)$ and $t(\tau)$, evolving in the evolution parameter $\tau$, and governed by the action
\be
  S=\frac{m}{2} \int d\tau \left(\frac{\dot q^2}{\dot t}-\omega^2 \dot t\, q^2 \right)
\ee
where the dot indicate the derivative with respect to $\tau$.  It is immediate to see that this is physically fully equivalent to the harmonic oscillator discussed in the previous section. In fact, the equation of motion for $q$ is
\be
  \frac{d}{d\tau}\frac{\dot q}{\dot t}=-\omega^2 \dot t \, q
  \label{uno}
\ee
which gives immediately the harmonic oscillator equation $d^2q/dt^2=-\omega^2q$; while the equation for $t$ is 
\be
  \frac{d}{d\tau}\left(\frac{\dot q^2}{\dot t^2}+\omega^2 q^2\right)=0,
  \label{due}
\ee
which is not an independent equation: it is simply the conservation of energy that follows from \eqref{uno}. The system has indeed a large gauge invariance, under arbitrary reparametrization of its independent variable $\tau$.  In this, it is very similar to general relativity, which is equally invariant under the reparametrization of its independent coordinate variables (\emph{Diff}--invariance).

Let us discretize this system. As before, fix an interval in $\tau$, split it into $N$ steps of size $a$ and define $\tau_n=na$, $t_n=t(\tau_n)$ and $q_n=q(\tau_n)$. Consider the discretized action 
\be
  S_N=\frac{m}{2} \sum_n a \left(\frac{(\frac{q_{n+1}-q_{n}}a)^2}{\frac{t_{n+1}-t_{n}}a}-\omega^2 \frac{t_{n+1}-t_{n}}a\, q_n^2 \right).
\label{actiondis}
\ee
Notice something important: the quantity $a$ drops from this expression. Indeed, the above reads
\be
  S_N=\frac{m}{2} \sum_n \left(\frac{({q_{n+1}-q_{n}})^2}{{t_{n+1}-t_{n}}}-\omega^2 (t_{n+1}-t_{n})\, q_n^2 \right).
\ee
In words, the discretized action is fully independent from $a$. This elementary observation is the main point of this article. Let us study the consequences of this fact.

The main consequence is that the continuum limit of the theory is not given by the double limit $N\to\infty,a\to 0$, but rather from the single limit $N\to\infty$. Let us see this more in detail.  

Define as before dimensionless variables $Q_n=\sqrt{\frac{m\omega}{\hbar} }q_n$ and $T_n=\omega t_n$. Notice that these are not defined using the time step $a$, which would be useless in this context since $a$ is not in the action, but rather the natural units given by the dynamics itself (in general relativity these natural units are provided by the Planck length). This yields the dimensionless action 
\begin{multline}
  \frac{S_N}{\hbar}=\frac{1}{2} \sum_n \left(\frac{({Q_{n+1}-Q_{n}})^2}{{T_{n+1}-T_{n}}}-(T_{n+1}-T_{n})Q_n^2 \right)\\ \equiv S_N(Q_n,T_n)
\end{multline}
Notice that the frequency has been absorbed in the normalization of the dimensionless variables. Suppose now we want to compute the same transition amplitude as before 
\be
W(q_f,t_f;q_i,t_i)= \int_{\substack{q(0)=q_i\\t(0)=t_i}}^{\substack{{q(1)=q_f}\\t(1)=t_f}}D[q(\tau)] D[t(\tau)]\ e^{\frac{i}{\hbar}{S[q(\tau),t(\tau)]}}.
 \ee
Then this is given by 
\be
W(q_f,t_f;q_i,t_i)= \lim_{N\to\infty}  \int dQ_n\, dT_n\ e^{i S_N(Q_n,T_n)}
\label{prop2}
 \ee
where $Q_0=\sqrt{\frac{\omega m}{\hbar} }q_i$ and $Q_N=\sqrt{\frac{\omega m}{\hbar} }q_f$,  
$T_0=\omega t_i$ and $T_N={{\omega} }t_f$.  There is no other limit to take than $N\to\infty$. When $N$ is large, the average time steps are automatically small. 

Below I study whether the classical and the quantum dynamics given by such discretization of the parametrized theory are well defined and sensible. 

\section{Classical dynamics} 

The standard discretization \eqref{discrete1} of the system breaks the interval of the physical time $t$ into $N$ steps of equal size. The discrete equation of motion, obtained minimizing \eqref{discrete1} with respect to $q_n$ is 
\be
       v_{n+1}=v_{n}-a\;\omega^2 q_n.
       \label{unod}
\ee
where I have defined the discrete velocity
\be
v_{n+1}\equiv \frac{q_{n+1}-q_{n}}{a}
\ee
Equation \eqref{unod} gives the velocity at the next time-step in terms of the velocity at the previous  step and of the discrete impulse (the impulse is the force $-\omega^2 q_n$ times the time step $a$).  As well known, because of the approximation involved in the discretization, the energy 
\be
      E_n= \frac{m}{2}\left(v_{n+1}^2+\omega^2 q_n^2\right)
\ee
is not conserved in general.  

Consider now the discretization of the parametrized system.  The key observation is that while the two continuous equations of motion \eqref{uno} and \eqref{due} are degenerate (the second follow from the first), their discretization {\em are independent equations}.  The first discretized equation is  as before
\be
        v_{n+1}=v_{n}-(t_{n+1}-t_{n})\; \omega^2 q_n.
               \label{unodp}
\ee
where the discrete velocity is now
\be
v_{n+1}\equiv \frac{q_{n+1}-q_{n}}{t_{n+1}-t_{n}}
\ee
and the fixed size time step $a$ is replaced by the variable size time step $(t_{n+1}-t_{n})$. What about the second equation? The variation of the action with respect to $t_n$ gives easily
\be
      E_{n+1}= E_{n}.
               \label{duedp}
\ee
That is, energy is now conserved!  How is it possible?   It is possible because the additional degree of freedom, which is the position of the time steps, is now adjusted by the dynamics in order for the energy to be conserved.  In other words, the \emph{continous} parametrized system adds a degree of freedom which is fully gauge. The \emph{discrete} non-parametrized dynamics breaks energy conservation.  But the \emph{discrete parametrized}  dynamics breaks the gauge freedom added by the parametrizations and exploits it by fixing it so that energy is conserved.  Concretely, the ``positions" of the intermediate time steps $t_n$ are not gauge in the discrete theory: they are determined in order to adjust the conservation of energy. 

\begin{figure}
\centerline{\includegraphics[scale=.4]{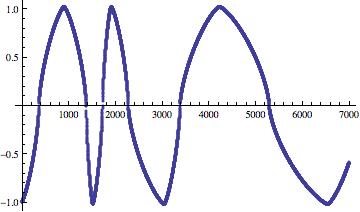}}
\caption{$x_n$ as a function of $n$ from the numerical integration of equations \eqref{unodp} and  \eqref{duedp}.}
\vspace{4mm}
\centerline{\includegraphics[scale=.4]{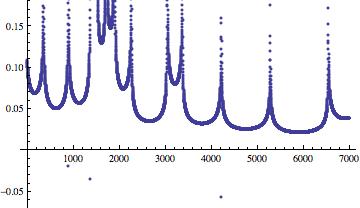}}
\caption{The size of the time step $(t_{n+1}-t_n)$ as a function of $n$ during the integration of Figure 1.}
\vspace{4mm}
\centerline{\includegraphics[scale=.4]{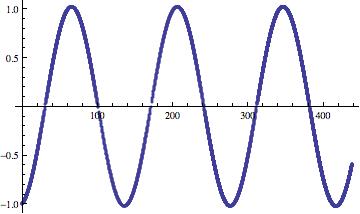}}
\caption{A two dimensional plot of $x_n,t_n,$ from the numerical integration of equations \eqref{unodp} and  \eqref{duedp}.}
\end{figure}

The integration of the discrete equations of motion of the parametrized systems can be performed numerically, showing that they give the correct dynamics. See Figures 1, 2 and 3.  Notice the irregular evolution of $q_n$ (Figure 1) and $t_n$ (Figure 2) as functions of $n$: This reflects the arbitrary dependence on $\tau$ of the parametrized system, gauge-fixed in the discretization (and determined by the initial data). But then $q_n$ and $t_n$ combine into the well-known solution of the harmonic oscillator equation, in their relative evolution (Figure 3).

The difference between the discretization of a standard system and that of a parametrized system can be directly seen in the equations of motion.  In the first case, the equations of motion \eqref{unod} contain explicitly the discretization size $a$. Therefore we are in fact dealing with a one-parameter family of equations, which converge to the continuum theory when the limit $a\to0$ is appropriately taken. Specifically, if $q_n(q_0,v_0,a)$ is the solution of the equations with initial values $q_0$ and $q_1=q_0+av_0$, then the solution of the continuum theory is recovered as
\be
    q(t) = \lim_{a\to0}\ q_{\frac{t}{a}}(q_0,v_0,a).
\ee
In the parametrized case, instead, {\em the discrete equations of motion (\ref{unodp},\ref{duedp}) do not contain $a$}. The continuum theory is not recovered by tuning a parameter in the equation, but rather by choosing initial (or boundary) data appropriately. Specifically, if $q_n(q_0,t_0,v_0,\Delta t_0)$ and  $t_n(q_0,t_0,v_0,\Delta t_0)$ give the solution with initial data $(q_0,t_0,q_1=q_0+\Delta t_0 v_0,t_1=t_0+\Delta t_0)$, then we can define $n(t)=n_t(q_0,t_0,v_0,\Delta t_0)$ as the inverse of $t=t_n$ (defined for appropriate values of $t$) and the continuum limit is given by 
\be
    q(t) = \lim_{\Delta t\to0}\ q_{n_t(q_0,t_0,v_0,\Delta t_0)}(q_0,t_0,v_0,\Delta t_0).
\ee
That is, the limit is not taken by tuning a constant, but rather by sending the initial data to an appropriate limit. 

This difference becomes far more clear, and physically more interesting, if instead of fixing a solution by means of initial data (initial position $q_0$ and velocity $v_0$ at some time $t_0$), we fix it by giving boundary data (initial and final positions ($q_i,q_f$) at some time given interval $[t_i, t_f]$).  Say we fix the number of discretization steps to be $N$.  Let $q_n(q_i,t_i;q_f,t_f;a)$ be the solution of the \eqref{unod} with 
boundary data $q_0=q_i$,  $t_0=t_i$,  $q_{N+1}=q_f$,  $t_{N+1}=t_f$.  Then the continuous solution is recovered as the limit 
\be
    q(t) = \lim_{\substack{a\to0}{N\to\infty}}\ q_{\frac{t}{a}}(q_i,t_i;q_f,t_f;a).
\ee
while in the parametrized case we have just 
\be
    q(t) = \lim_{N\to0}\ q_{n_t(q_i,t_i;q_f,t_f;)}(q_i,t_i;q_f,t_f;).
\ee
because increasing the number of steps at fixed boundary times reduces automatically the time steps to zero.

Before concluding this section, we remark that, as noticed by  Dittrich and Bahr in \cite{Bahr:2009qc}, all this does not happen if instead of discretizing the action as in \eqref{actiondis}, we choose a very special discretization. This special choice is called the ``perfect action" by Dittrich and Bahr, and is obtained by choosing the action of each time step to be the Hamilton function of the theory, namely the value of the action on the physical trajectory, expressed as a function of the initial and final values of the variables.  For the harmonic oscillator the Hamilton function is 
\begin{multline}
S(q,t;q',t')={\omega m}\frac{({q'}^2+q^2)\cos{\omega(t'-t)}-2q'q}{2\sin{\omega(t'-t)}}
\end{multline}
which has the property 
\be
S(q,t;q',t')+S(q',t';q'',t'')=S(q,t;q',t')
\ee
when $q',t'$ is the solution of the equations of motion with initial and final values $q,t,q'',t''$.
The perfect discretization is therefore
\be
S=\sum_n S(q_{n+1},t_{n+1};q_{n},t_{n}).
\ee
This has the property of being independent from $N$, once initial and final values are specified, and to conserve the full reparametrization invariance of the continuous theory.  The possibility of using the perfect action or approximations of the same in quantum gravity have been investigated by Dittrich. Here instead I focus on the possibility of using a simple discretization, and exploiting precisely the fact that it fixes the reparametrization invariance gauge. 

\section{Quantum dynamics} 

Consider the quantum theory.  The first important observation is that  quantities that make sense in the discretization of the unparametrized theory do not make sense in the parametrized one.   For instance, in the unparametrized theory we can consider the two point function 
\be
W(k) = \langle q_k q_0 \rangle = \int dq_n\  q_k q_0\ e^{\frac{i}{\hbar}S(q_n)}
\ee
The physical quantity 
\be
W(t) = \langle 0 |q(t) q(0)|0 \rangle= \langle 0 |q e^{-iHt} q|0 \rangle
\ee
can be obtained from $W(k)$ by taking the limit $k\to\infty, \Omega\to 0$ with $t=ka$. In the parametrized case,  the corresponding quantity
$W(\tau)$ would make no sense, since $\tau$ has no physical meaning.  The quantity
\be
W(\tau) = \langle 0 |q(\tau) q(0)|0 \rangle= \langle 0 |q e^{-iH\tau} q|0 \rangle
\ee
where here $H$ is the generator of the evolution in $\tau$, is indeed independent from $\tau$ (because of the reparametrization invariance, which gives $H=0$) and carry no physical information. The corresponding difficulties in quantum gravity are well known and widely discussed.

To extract information from the theory we must use instead the propagator \eqref{prop}, namely place the physical inputs of the calculation on the \emph{boundary} of a finite region.\footnote{The relation between 
$W(t)$ and $W(q,t;q',t')$ is then easily obtained from the explicit expression of the vacuum $\psi_0(q)=\langle q|0 \rangle$ which can be used to transform the two quantities into each other:
\begin{multline}
W(t)=\langle 0|q(t)q(0)|0  \rangle 
=\langle 0|\hat q e^{-iHt} \hat q|0  \rangle \\
=\int dq dq'\  \langle 0|q\rangle q \langle q| e^{-iHt}  |q' \rangle q'\langle q'|0  \rangle \\
= \int dq dq'\ W(q,t;q',t')\ q' q\  \overline{\psi_0(q)} \psi_0(q).
\end{multline}
In field theory, this technique is used to transform the field propagator into particle propagator. The last line of this equation is the definition of the boundary technique used  in loop quantum gravity to compute particle's $n$-point functions \cite{Modesto:2005sj,Rovelli:2005yj,Bianchi:2006uf,Alesci:2008ff,Alesci:2007tx,Alesci:2007tg,Bianchi:2011fk}.}
From \eqref{prop}, we have
 \be
W(q_f,t_f;q_i,t_i)
=\tilde W\left(\sqrt{\frac{\omega m}{\hbar}}q_f,\omega t_f;\sqrt{\frac{\omega m}{\hbar}}q_i,{{\omega}}t_i\right)
 \ee
where the propagator of the dimensionless quantities is 
 \be
\tilde W(Q_f,T_f;Q_i,T_i)
= \lim_{N\to\infty} {\cal N}  \int d\mu(Q_n,T_n)\ e^{i S_N(Q_n,T_n)}.
 \ee
 where $Q_{N+1}=Q_f$, $Q_0=Q_i$, $T_{N+1}=T$ and $T_{0}=T_i$. 
To fully define this quantity, we need to fix the domain of the integration and the integration measure. 

The domain of integration for $Q_n$ variables is clearly the entire real line.  For the $T_n$ variables, let us consider the restriction $T_i<T_n<T_{n+1}<T_f$. This is reasonable in order to have only ``forward" propagation in time.\footnote{Is there a relation between this condition and cell orientation in spinfoams?}  The integration measure can be obtained in various ways (see  \cite{Bahr:2011uj})). The simplest is to study the free case $\omega=0$, where the integration can be performed explicitly. A straightforward  calculation sketched in the next section shows that taking
\be
{\cal N}= \frac{N!}{(T_f-T_i)^N}
\label{m1}
\ee
and
\be
d\mu(Q_n,T_n)= 
 \frac{\prod_{n,1}^N dQ_n\,dT_n}{\prod_{n,0}^N\sqrt{2\pi(T_{n+1}-T_{n})}}
\label{m2}
\ee
we obtain the correct free particle propagator 
\be
W^{0}(q_f,q_i,t)=\sqrt{\frac{m}{2\pi i\hbar t}}\ e^{-i\frac{m(q_f-q_i)^2}{2\hbar t}}
\ee
when $\omega=0$.  It is therefore natural to try the ansatz of fixing the integration measure by (\ref{m1},\ref{m2}).  With this, the discretized path integral is fully defined. 

I have computed the discretized path integral numerically in the Euclidean regime, for $\omega\ne0$. The result converges nicely to the (Euclidean continuation of the) well known exact result. 
\be
W(q_f,q_i,t)=\sqrt{\frac{\omega m}{2\pi\hbar\sinh{\omega t}}}\ e^{-\omega m\frac{(q_f^2+q_i^2)\cosh{\omega t}-2q_fq_i}{2\hbar\sinh{\omega t}}}.
\label{exact}
\ee
See Figure 4 for an exemple of the result of the numerical integration. 

\begin{figure}
\centerline{\includegraphics[scale=.7]{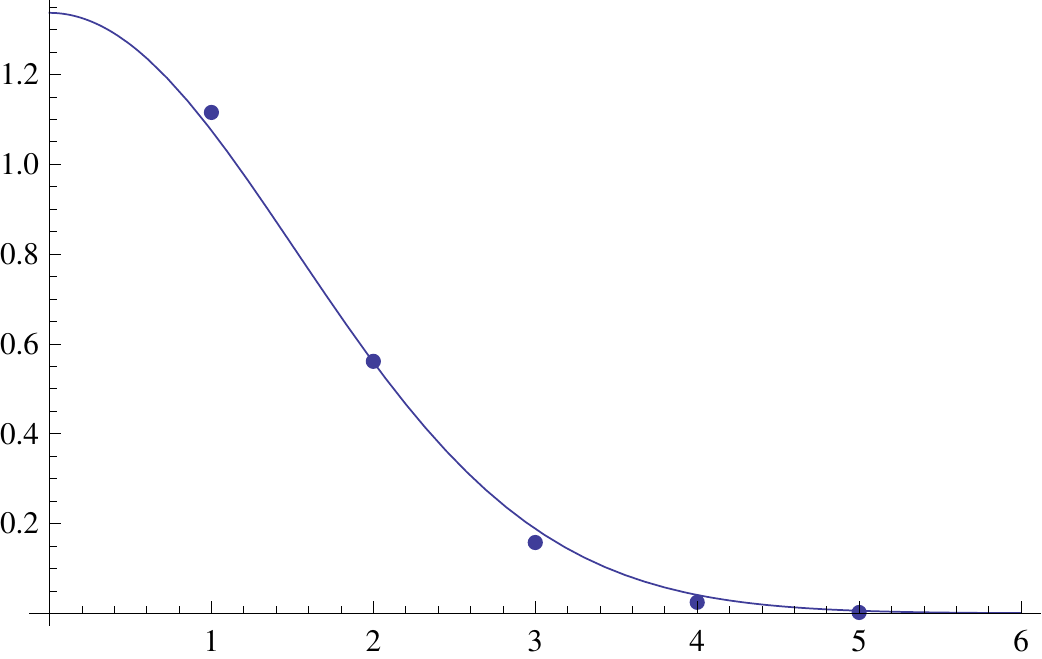}}
\caption{Numerical integration of the discretized path integral of the parametrized oscillator. The graph gives 
the Euclidean evaluation of $W(q_f,t_f;q_i,t_i)$ as a function of $q_f$, with $\omega t\sim1$, $t_i=0$ and $q_i=0$, compared with the exact result \eqref{exact}. The number of integration points is $N= 2$.}
\end{figure}

\section{\emph{Ditt}--invariance}

When the potential is negligible with respect to the kinetic term, the discretized equations of motion  (\ref{unodp},\ref{duedp}) become
\be
        v_{n+1} \simeq v_{n}
               \label{unodp2}
\ee
and 
\be
\frac{v^2_{n+1}}2 \simeq \frac{v^2_{n}}2.
                       \label{duedp2}
\ee
That is, the second equation is again dependent on the first, as in the continuum parametrized theory. Therefore the invariance under diffeomorphisms on the parameter time $\tau$ (\emph{Diff}--invariance), broken by the discretization, is recovered within the discretized theory in this regime. In terms of boundary values, this regime is approached when 
\be
       {\omega ({t_f-t_i})}\ll {\left| \frac{q_f-q_i}{q_i}  \right|} 
\label{flat}
\ee
In this regime the classical trajectory  is well approximated by a straight line, namely a trajectory with no curvature. Such approximate recovery of \emph{Diff}--invariance near the ``flat" trajectories strongly recalls the recovery of \emph{Diff}--invariance of Regge calculus near flat space studied by Bianca Dittrich. It seems to be a general phenomenon for the discretization of  reparametrization-invariant invariant systems.  Let us see what are its consequences on the discretized path integral. 

In the limit in which the potential term of the action can be disregarded, the discretized path integral can be performed explicitly. Repeated use of
\be
        \int dq\ e^{-\frac{(a-q)^2}{2t_1}-\frac{(b-q)^2}{2t_2}}=\sqrt{\frac{2\pi\ t_1t_2}{t_1+t_2}}e^{-\frac{(a-b)^2}{2(t_1+t_2)}}
\ee
gives
\be
        \int dq_n\ e^{-\sum_n\frac{(q_{n+1}-q_n)^2}{2t_n}}=(2\pi)^{\frac{N}{2}}\sqrt{\frac{\prod_k t_k}{\sum_k t_k}}e^{-\frac{(q_{n+1}-q_n)^2}{2\sum_k t_k}}
\ee
where the sums in $n$ go from 1 to $N$ while the sums in $k$ go from 1 to $N+1$. Fixing $\sum_k t_k=t$ and using
\begin{multline}
        \int_0^{T_{N+1}} \hspace{-5mm}dt_N\int_0^{t_N} \hspace{-3mm}dt_{N-1}... \int_0^{t_2} \hspace{-3mm}dt_{1}
        \ \  \sqrt{{\prod_{k} (t_{k}-t_{k-1})}}=\\
         \frac{(T_{N+1}-T_{0})^N}{N!}
\end{multline}
We obtain that the discretized path integral in the limit of vanishing $\omega$ with $N$ steps
\be
 {\cal N}  \int d\mu(Q_n,T_n)\ e^{-\frac{1}{2} \sum_n \frac{({Q_{n+1}-Q_{n}})^2}{{T_{n+1}-T_{n}}}}.
\ee
is actually independent from $N$. This is the  \emph{Ditt}--invariance of the functional integral. 

A key consequence of this is that the expansion for small number $N$ of steps is very good in the ``flat" regime \eqref{flat}. This can be checked numerically. See for instance Figure 4, where the exact transition amplitude is obtained with an approximation of a few \% simply with $N=3$. Even $N=1$ gives a very good approximation of the exact transition amplitude when sufficiently near flatness.

In other words, the discretization of the parametrized systems behaves like perturbation theory. The expansion in the number of steps $N$ is a good expansion in the regime \eqref{flat}.

\section{Conclusion}

The system studied here is too simple  to allow deriving general conclusions from it. Nevertheless, the analysis appears definitely to suggest that the classical and quantum discretization of  \emph{Diff}--invariant systems behaves quite differently from that of standard systems.  

The most remarkable feature of these systems is that the continuum limit is obtained directly taking the number of steps to infinity, without tuning a parameter in the action to a critical value.  This changes drastically the structure of the continuum limit from-well studied cases such as lattice QCD.\footnote{To avoid possible misunderstanding, let me observe that the relevant distinction here is not gravity versus strong interactions: it is whether the action being discretized is invariant under change of independent variables (that is, reparametrization invariant, or diff invariant, or invariant under general coordinate transformations) or not. This should be clear from the example studied, where the two actions contrasted refer to the same physics. The relation with gravity is that when we neglect gravity the simplest form of the action describing the real would is background dependent; while when we include gravity, it is reparamertrization invariant.}

Furthermore, the systems admits a regime where the approximation of the transition amplitudes is very good already for a very small number of integration steps.   For the system studied in this paper, this is the regime \eqref{flat}, where the classical trajectories approach flatness. 

These same two phenomena have appeared in quantum gravity. First, the continuous amplitudes appear to be given simply by taking the number of cells, or the two-complex, to infinity, without the need of tuning a parameter to its critical value \cite{Rovelli:2010qx, Rovelli:2011eq}. Second, the ``vertex expansion" of the transition amplitudes appear to give excellent agreement with the expected value, even at $N=1$ \cite{Bianchi:2009ri,Rovelli:2011kf,Krajewski:2010yq,Bonzom:2011br}. These unexpected phenomena appeared difficult to understand in quantum gravity. In particular the implicit expansion around flat space appeared to be problematic because a gauge freedom is there for the linearized theory, but is broken to higher order \cite{Dittrich:2009fb}.  The example shown clarifies what happens, makes the origin of these phenomena transparent,  and shows that \emph{Ditt}--invariance, far from causing problems, is in fact the source of the magic that makes the expansion viable.

To be sure, the analogy does not need to hold necessarily.  Field theoretical aspects of the problem, and in particular radiative corrections, might significantly change  the situation in quantum gravity \cite{Crane:2001as,Perini:2008pd,Geloun:2010vj}.  However, as recalled in Section 2, notice that the continuum limit of a conventional discretized system is obtained tuning a parameter also for finite dimensional systems: therefore, the tuning of the parameter is \emph{not} a field theoretical effect.  The fact that the tuning is not required for a parametrized system, even in a one-dimensional case, is therefore significative for quantum gravity. 

In the Appendix below I discuss in which sense the discretized theory preserves  \emph{Diff}--invariance, and in which sense  \emph{Diff}--invariance is broken.  The breaking of diff-invariance of the discretized theory has been pointed out repeatedly \cite{Gambini:2008as,Bahr:2009mc,Bahr:2011uj}. Such breaking of  \emph{Diff}--invariance, however, is \emph{not} a problem for the theory, because the exact transition amplitudes are the ones in the limit, not the ones for fixed $N$ (or fixed foam), and these are ok, as the example in this paper shows. In the Appendix, I argue in detail why it is not a difficulty in quantum gravity either.

Finally, an intriguing aspect of the issue is the appearance of the  ``topological" flat phase, where the amplitudes are independent from the number of points of the discretization.  It is tempting to speculate that the same phenomenon happens in general relativity, with the ``topological" flat phase being given by BF theory.  The suggestion that quantum gravity transition amplitudes could approach BF theory (where the connection is flat) in some regime has been made repeatedly, and \emph{Ditt}--invariance provides a concrete mechanism for this to happen.  

Intuitively, when increasing the number of discretization points $N$ keeping the time interval $T_f-T_i$ fixed, each  time step in the path integral becomes very small in average. But having very small time-steps means being deep into the regime of \emph{Ditt}--invariance, and in this regime any further increase of $N$ does not change the amplitude.  This is what leads to the convergence in $N\to\infty$.  Does the same happen in quantum gravity?

\centerline{---}
Thanks to Alberto Ramos and Laurent Lellouch for lectures and conversations on the subject. Thanks to Bianca Dittrich for numerous exchanges and an accurate reading of the first version of this article.

\appendix
\section*{Appendix: Does discretization break \emph{Diff} invariance?}

I have stated in the main text that the discretization breaks \emph{Diff}--invariance of the parametrized system.  This statement appears to contradict the idea that a discretized theory is still \emph{Diff}--invariant, an idea that has been repeatedly defended, for instance by Tullio Regge himself in his introduction of the Regge discretization of general relativity \cite{Regge:1961px}.  Here I show that the two points of view are not  in contradiction with one another. They only use a different language. 

Let me explain in which sense discretization does \emph{not} break \emph{Diff}--invariance.  Fix the initial and final times and positions $(q_i,t_i;q_f,t_f)$.  The continuous equations of motion are defined for a curve $q:[t_i,t_f]\to R$. Call  ${\cal Q}$ the space of these curves (say, to be more definite: continuous and almost everywhere twice differentiable).  The solution of the equation, generically, will be a curve $q\in{\cal Q}$.  If we discretize the interval with $N$ steps, the solution of \eqref{unod} determines a piecewise linear curve $q_{N}\in{\cal Q}$, which generically converges pointwise to $q$ as $N\to\infty$.

Consider next the continuous parametrized system. The equations of motions are now for the two functions  $q:[\tau_i,\tau_f]\to R$ and $t:[\tau_i,\tau_f]\to R$ (with $dt/d\tau>0$).   Let $\cal G$ be the space of such curves.  Given $(\tau_i,q_i,t_i;,\tau_f,q_f,t_f)$ the equations of motion (\ref{uno},\ref{due}) \emph{do not} determine a unique solution, because of \emph{Diff}--invariance.  There is a projection from $\cal G$ to $\cal Q$, given (in physicists' notation) by $(q(\tau),t(\tau))\mapsto q(t)=q(\tau(t))$ which sends  gauge equivalent solutions into the same physical trajectory.  Up to this gauge, the solution is unique. Furthermore, the initial and final values $(\tau_i,\tau_f)$ become irrelevant under such gauge invariance. That is, solutions with different $\tau$ boundary values are gauge equivalent. 

Finally, let's come to the system of interest, which is the discretization of the parametrized system. Choose as before the number $N$ of steps of the discretization.  Given $(\tau_i,\tau_f,q_i,t_i;q_f,t_f)$ the equations of motion (\ref{unodp},\ref{duedp}) \emph{do} determine now a unique solution, since they are not anymore independent. This is the ``breaking"  of \emph{Diff}--invariance.   However, it is still true that the solution does \emph{not} depend on the boundary values $(\tau_i,\tau_f)$, in the sense that the projection of the solution in $\cal Q$ is independent from  $\tau_i$ and $\tau_f$.   Therefore there is still only one physical distinct solution for each boundary set  $(q_i,t_i;q_f,t_f)$. The number of physical solutions (that is, the dimension of the phase space), has not increased.

More importantly, the solution is now given by the sequence $(q_n,t_n), n=1,...,N$. At first sight, this much resembles the continuum expression $(q(\tau), t(\tau))$, but there is a crucial difference: $(q(\tau), t(\tau))$ identifies a curve in $\cal Q$ only with a large redundancy, that is, different couples of function $(q(\tau), t(\tau))$ determines the same curve $q(t)$. But not so for $(q_n,t_n)$. The map from the sequences $(q_n,t_n)$ to $\cal Q$ is injective.  This follows from the fact that, generically, given the image of $(q_n,t_n)$ in $\cal Q$, we can easily reconstruct the sequence $(q_n,t_n)$: this is given by the times $t_n$ where the curve fails to be straight, and the corresponding values $q_n=q(t_n)$ of the position.  This inversion can only fail if there is an $n$ where there is no curvature, that is, $\frac{q_{n+1}-q_{n}}{t_{n+1}-t_{n}}=\frac{q_{n}-q_{n-1}}{t_{n}-t_{n-1}}$, but this is generically not allowed by the equations of motion, except in the large $N$ limit itself, which is where \emph{Ditt}--invariance manifests itself. 

In other words, the space of the solutions of the discrete parametrized equation is formed generically by curves that are genuinely distinct when projected to $\cal Q$.  In this sense, the discretized system does not deal with trajectories defined up to gauge, but directly with gauge invariant 
trajectories. 

 In the context of general relativity, this translates into the fact that Regge metrics can be interpreted as genuine geometries (that is, equivalent classes of metrics under diffeomorphisms).   Indeed, let $\cal Q$ be the space of the 4d riemannian geometries (say continuous and twice differentiable almost everywhere).  Let $\cal G$ be the space of the 4d metric tensors.  A projection from $\cal G$ to $\cal Q$ is obtained identifying any two metrics related by a diffeomorphism.    Now, the space $\cal R$ of the Regge geometries can be identified as the subspace of $\cal G$ formed by the continuous geometries that are flat almost everywhere, except on the two skeleton of a triangulation immersed in the space.  In general, we do not know how to cohordinatize the space of the geometries $\cal G$, but Regge has found a remarkable way to cohordinatize its subspace $\cal R$: a point in $\cal R$ is  determined by the physical length $l_e$ of the edges $e$ of the cellular decomposition.  The quantities $l_e$ are the Regge variables.  They cohordinatize physical geometries, and are fully coordinate independent. They are invariant under diffeomorphisms.  In other words, the Regge lengths $l_e$ are not distances between arbitrary coordinate points: they are the physical lengths of the sides of the triangles where the geometry fails to be flat. A such they are cohordinate independent quantities. 
 
This is the sense in which Tullio Regge asserted (correctly) that Regge calculus is a \emph{Diff}--invariant way of treating general relativity. And this is the sense in which the discretized equations (\ref{unodp},\ref{duedp}), or the spinfoam amplitudes on a given foam,  are gauge invariant. 

On the other hand, the discretized theory is \emph{not} \emph{Diff}--invariant in the following sense.  Fix an $N$ and a discretization scale.  Let $(q(\tau),t(\tau))$ be a solution of the continuous equation of motion and $(\tilde q(\tau)=q(f(\tau)),\tilde t(\tau)=t(f(\tau)))$ a gauge equivalent solution. Then generically $(q_n=q(\tau_n),t_n=t(\tau_n))$ and $(\tilde q_n=\tilde q(\tau_n), \tilde t_n=\tilde t(\tau_n))$ are not gauge equivalent in the discretized theory. Such breaking of diff-invariance has been pointed out repeatedly \cite{Gambini:2008as,Bahr:2009mc,Bahr:2011uj}. 

Such ``breaking of Diff invariance", however, is \emph{not} a problem for the theory, because the exact transition amplitudes are the ones in the limit, not the ones for fixed $N$ (or fixed foam), and these are ok, as the example in this paper shows.

One might object that in quantum gravity the discreteness is physical, because physical space is discrete, and the theory must be discrete as well as diff-invariant.  But this objection is based on a misunderstanding, because it confuses ``Planck scale discreteness of space" with the triangulation discretization.   The physical ``discreteness of space" is not given by the fact that we use discretization in the theory.  It is given by the fact that the areas and volumes
of these simplices take only discrete values.   The Planck-scale  discreteness is in the \emph{size} of the simplices, which takes only discrete values.

To clarify with an analogy: an electromagnetic field in a box can be expanded in discrete modes.  This is not quantum discreteness, of  course.  Quantum discreteness is that the energy of each mode is a multiple of $h\nu$.   If we truncate the theory by only taking a finite set of modes, then this is a truncation of degrees of freedom, nothing to do with quantum discreteness.  Here:  Fourier modes $\to$ variables on simplices; truncation to a finite number of modes $\to$ finite triangulation;  quantization of the energy $\to$ quantization of the size of each simplex (=Planck discreetness of space). A space-time described by few large simplices is analog to an electromagnetic wave formed by few large-wavelength modes with large amplitude; therefore large space times do not require fine triangulations to be effectively described, as suspected early \cite{Ashtekar:1992tm,Iwasaki:1992qy}.

So, the use of simplices in the theory is a discretization like the one in QCD, it is a truncation of the degrees of freedom. It can equally be done in the classical theory, and is not related to Planck scale discreteness.  Therefore the fact that diff invariance is broken in this sense on a finite truncation is not a difficuty for the theory.

\vfill
\vfill

\bibliographystyle{apsrev4-1}
\bibliography{BiblioCarlo}
\end{document}